\documentclass[a4paper]{article}

\usepackage{multirow}

\usepackage{INTERSPEECH2020}

\title{Ultra Fast Speech Separation Model with Teacher Student Learning}

\name{\begin{tabular}{c}Sanyuan Chen$^{\ddagger}$, Yu Wu$^{\dagger}$,  Zhuo Chen$^{\dagger}$, Jian Wu$^{\dagger}$, Takuya Yoshioka$^{\dagger}$ \\ Shujie Liu$^{\dagger}$, Jinyu Li$^{\dagger}$, Xiangzhan Yu$^{\ddagger}$\end{tabular}}
\address{
$^{\ddagger}$Harbin Institute of Technology, China \\
$^{\dagger}$Microsoft Corporation, USA 
	}
	
\email{ 
\begin{tabular}{c} 
sychen@ir.hit.edu.cn \\ \{yuwu1, zhuc, wujian, tayoshio, shujliu, jinyli\}@microsoft.com,  yxz@hit.edu.cn 
\end{tabular}
}

\begin{document}

\maketitle
\begin{abstract}
Transformer has been successfully applied to speech separation recently with its strong long-dependency modeling capacity using a self-attention mechanism. However, Transformer tends to have heavy run-time costs due to the deep encoder layers, which hinders its deployment on edge devices. A small Transformer model with fewer encoder layers is preferred for computational efficiency, but it is prone to performance degradation. In this paper, an ultra fast speech separation Transformer model is proposed to achieve both better performance and efficiency with teacher student learning (T-S learning). We introduce layer-wise T-S learning and objective shifting mechanisms to guide the small student model to learn intermediate representations from the large teacher model. Compared with the small Transformer model trained from scratch, the proposed T-S learning method reduces the word error rate (WER) by more than 5$\%$ for both multi-channel and single-channel speech separation on LibriCSS dataset. Utilizing more unlabeled speech data, our ultra fast speech separation models achieve more than 10$\%$ relative WER reduction.

\end{abstract}
\noindent\textbf{Index Terms}: speech separation, Teacher Student Learning, Transformer, deep learning

\section{Introduction}

Speech separation plays a vital role in front-end speech processing, aiming to handle the cocktail party problem.  
Recently, with the success of Transformer model in speech community \cite{dong2018speech, Li2020Comparison}, the Transformer \cite{chang2020end, chen2020dual} and its variants \cite{chen2020continuous2} have successfully achieved superior performance on this task.
However, these models tend to have heavy run-time costs due to the deep encoder layers, while the real-time inference is crucial for product deployment especially on resource limited edge devices.

Given the great demand of better computational efficiency, a small speech separation model is preferred for the deployment, with considerably fewer encoder layers and fast inference speed.
Unfortunately, the use of the smaller model directly tends to degrade the separation performance and thus hurts performance of downstream tasks such as multi-speaker speech recognition \cite{chen2020continuous2}.

To build a small model with both fast inference speed while maintaining the accuracy, teacher student learning (T-S learning) is a common strategy for model training, and has been shown effective in various tasks \cite{li2014learning, hinton2015distilling}.
With the T-S learning, a smaller Transformer based separation model (student) is trained to mimic the behavior of a large pretrained model (teacher). In this work, we apply the T-S learning to fast transformer based separation network training, and introduce three updates to further enhance the performance.
Specifically, with the help of \textit{Layer-wise T-S learning}, not only the final prediction but also the intermediate feature maps of the teacher model are leveraged.
Since the teacher model is not perfect and may generate results with noises and errors, we introduce an \textit{Objective Shifting} mechanism to let the learning objective gradually shift from the teacher predictions to the golden predictions. 
Going beyond the limitation of the labelled training data, large-scale \textit{unlabeled speech separation data} are used in our T-S learning, to allow the student to better capture teacher's behaviours. 
Different from previous work applying T-S learning for speech enhancement and separation, which train student models in the same model size while operating at different input features \cite{watanabe2017student, subramanian2018student}, this paper aims to distill the teacher model's knowledge to create a smaller and faster student model. Besides, this paper is the first one to use a large amount of unlabeled data in the T-S learning for speech separation. 

We conduct the experiment on the public LibriCSS dataset \cite{chen2020continuous}. The experimental results show that our ultra fast Transformer model can achieve more than 5\% average relative WER gains with our proposed T-S learning for both single-channel and multi-channel speech separation, and the improvements are more significant for the utterances with higher overlap ratio. 
Several ablation experiments show that both Layer-wise T-S learning and Objective Shifting mechanisms are crucial to the performance improvements.
Moreover, since annotated data are not required for the Layer-wise T-S learning, pretraining on large-scale unlabeled data enables our ultra fast Transformer model achieve more than 10\% average relative WER gains with the proposed T-S learning methods.

\section{Background}

\subsection{Problem Formulation}
Continuous speech separation (CSS) aims to estimate individual speaker signals from a continuous speech input where the source signals are fully or partially overlapped.
Let $y(t)$ denote the mixed signal and $x_s(t)$ the $s$-th individual target signal, where $t$ is the time index. The mixed signal is modeled as follows: 
\begin{equation}
    y(t)=\sum_{s=1}^S x_s(t). 
\end{equation} 
Their short-time Fourier transforms (STFTs) are denoted as $\mathbf{Y}(t,f)$ and $\mathbf{X}_s(t,f)$,  respectively. $f$ denotes frequency index.

Following \cite{wang2014training, erdogan2017deep}, instead of directly outputting the STFT of the individual signals $[\mathbf{X}_1(t,f) \ldots \mathbf{X}_S(t,f)]$, we employ the mask learning to recover the clean speech, where a group of masks $\mathbf{M}(t,f) = [\mathbf{M}_1(t,f) \ldots \mathbf{M}_S(t,f)]$ are firstly estimated with a deep learning model $F(\cdot)$.  
Then, for the $s$-th individual signal, $\mathbf{X}_s(t,f)$ is obtained either by mask-based beamforming or by direct masking, i.e., $\mathbf{M}_s(t,f) \odot  \mathbf{Y}^1(t,f)$ where $\odot$ is the element-wise product, $\mathbf{Y}^1(t,f)$ is the first  channel of $\mathbf{Y}(t,f)$. 

\subsection{Transformer Model}

\begin{figure}[t]		
	\begin{center}
		\includegraphics[width=0.8\columnwidth]{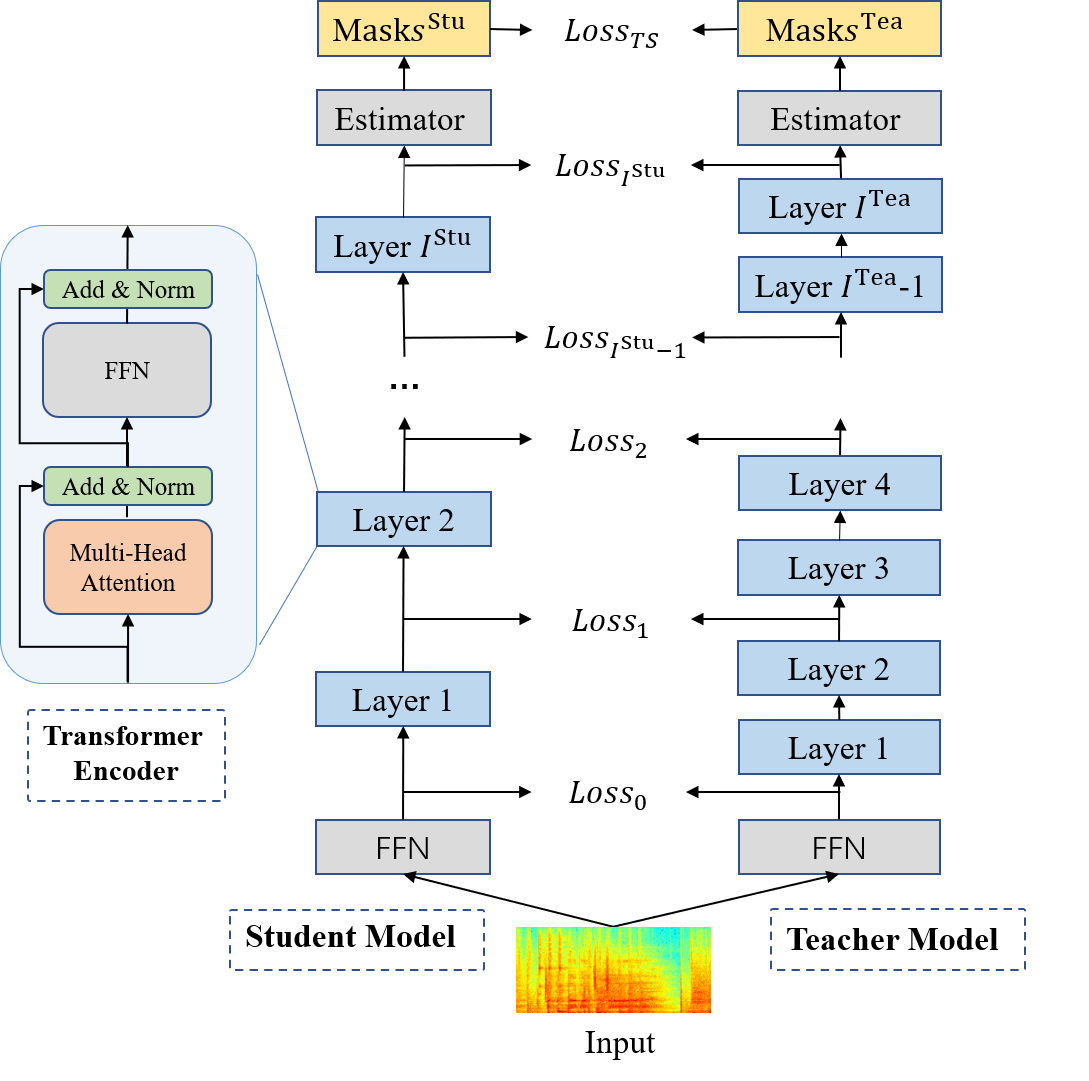}
	\end{center}
	\vspace{-2em}
	
	\caption{Layer-wise Teacher Student Learning of Transformer model. 
	}\label{fig:kd_transformer} 
    \label{ssec:model}\vspace{-2em}
\end{figure}

As shown in the Figure~\ref{fig:kd_transformer}, The Transformer model \cite{vaswani2017attention} is composed of a stack of identical Transformer encoder layers, each of which consists of a multi-head self-attention module and a position-wise fully connected feed-forward module.

Before sending to Transformer encoder, for both single and multi channel separation network, the input feature $\mathbf{Y}(t,f)$ is projected to representation  $\mathbf{h}_{0}$  with fixed dimension, by a feed-forward module  $\text{FFN}(\cdot)$: 
\begin{flalign}
\mathbf{h}_{0} & =\text{FFN}(\mathbf{Y}(t,f)). 
\end{flalign}

Given the input, $\mathbf{h}_{i-1}$, of the $i$-th layer, the output $\mathbf{h}_{i}$ is calculated as
\begin{flalign}
\mathbf{h}_{i}' & =\text{layernorm} (\mathbf{h}_{i-1} + \text{MultiHeadAttention} (\mathbf{h}_{i-1}))  \\
\mathbf{h}_{i} &= \text{layernorm} (\mathbf{h}_{i}' + \text{FFN}(\mathbf{h}_{i}')), 
\end{flalign} where
$\text{MultiHeadAttention}(\cdot)$ and $\text{layernorm}(\cdot)$ denote the multi-head self-attention module and the layer normalization, respectively. 
The multi-head self-attention module is implemented with relative position embedding as \cite{shaw2018self, chen2020continuous2, chen2020don}.

Given $\mathbf{h}_{I}$, the output of the final layer, we obtain the masks $\mathbf{M}(t,f)$ with $\text{Estimator}(\cdot)$, an estimator consisting of a feed-forward module and a sigmoid activation function, i.e., 
\begin{flalign} 
    \mathbf{M}(t,f) &= \text{Estimator}(\mathbf{h}_{I}) \\
    &= \text{sigmoid}( \text{FFN}(\mathbf{h}_{I})). 
\end{flalign}

\subsection{Teacher Student Learning}

Teacher student learning is a common training strategy for model compression, where a smaller and faster student model is trained to generate the same output as a more powerful teacher model. Specifically, in separation task, the T-S learning can be represented as the minimization of the mean square error (MSE) between the estimated signals of the student and the teacher model:

\begin{flalign}
\mathcal{L}_\text{TS} = \frac{1}{T \times F \times S} \sum_{s=1}^S ||\mathbf{X}_s^\text{Stu}(t,f) - \mathbf{X}_s^\text{Tea}(t,f)||^2 
\end{flalign}
where $T$, $F$ and $S$ denote the number of the time frames, frequency bins and target signals, respectively. The estimated signals $\mathbf{X}_s^\text{Stu}(t,f)$ and $\mathbf{X}_s^\text{Tea}(t,f)$ is calculated as:
\begin{flalign}
\mathbf{X}_s^\text{Stu}(t,f) &= \mathbf{M}_s^\text{Stu}(t,f) \odot  \mathbf{Y}^1(t,f) \\ \mathbf{X}_s^\text{Tea}(t,f) &= \mathbf{M}_s^\text{Tea}(t,f) \odot  \mathbf{Y}^1(t,f)
\end{flalign}
where $\mathbf{M}_s^\text{Stu}(t,f)$ and $\mathbf{M}_s^\text{Tea}(t,f)$ are the estimated masks of the student model and the teacher model.

\section{Method}
To further enhance the efficacy of knowledge distillation, two mechanisms are introduced to baseline T-S learning, namely \textbf{Layer-wise T-S Learning} and \textbf{Objective Shifting}, that allows the student model to also benefit from Teacher's intermediate representation and oracle training label.
In addition, to further boost the performance of the student model, we leverage the \textbf{unlabeled data training} in our T-S learning framework.

\subsection{Layer-wise T-S Learning}

We introduce the layer-wise T-S Learning mechanism to train the student to reproduce not only the final prediction but also the intermediate outputs of the teacher model \cite{sun2019patient}.

As Figure~\ref{fig:kd_transformer} shows, given the $I^\text{Stu}$ layer student model  and $I^\text{Tea}$ layer teacher model,  
we minimize the mean square error (MSE) between the output of $i$-th layer of the student model and the corresponding $g(i)$-th output of the teacher model:
\begin{flalign}
\mathcal{L}_{i} = \frac{1}{T \times F} ||\mathbf{h}_{i}^\text{Stu} - \mathbf{h}_{g(i)}^\text{Tea}||^2 
\end{flalign} where $g(\cdot)$ is an uniform layer mapping function between indices from student layers to teacher layers.

Then the objective function of layer-wise T-S Learning is the weighted average function as:
\begin{flalign}
    \mathcal{L}_\text{LTS} = \frac{\sum_{i=0}^{I^\text{Stu}} (i+1) \cdot \mathcal{L}_i + (I^\text{Stu}+1) \cdot \mathcal{L}_\text{TS}}{\sum_{i=0}^{I^\text{Stu}} (i+1) + (I^\text{Stu}+1)}
\label{eq:lts} 
\end{flalign} where $\frac{i+1}{\sum_{i=0}^{I^\text{Stu}} (i+1) + (I^\text{Stu}+1)}$ is the weight for $\mathcal{L}_i$. The loss of a higher layer is assigned  with a larger weight  as \cite{chen2020don}. 

\subsection{Objective Shifting}

Since the student model is trained to recover the predictions of the teacher model, the performance of T-S learning would be limited to the teacher's capability.
To avoid this limitation, we introduce the Objective Shifting mechanism to train the student with both the teacher's prediction and training datasets \cite{kiperwasser2018scheduled, chen2020recall}.

Specifically, an additional loss item $\mathcal{L}_\text{PIT}$ in added to training objective, that minimizes the MSE between the estimated signals of the student model and the references in the training sets. 
The final loss function of layer-wise T-S learning with objective shifting is calculated as: 

\begin{flalign}
    \mathcal{L} &= \lambda(t) \mathcal{L}_\text{PIT} + (1-\lambda(t)) \mathcal{L}_\text{LTS} 
\label{eq:os}
\end{flalign} where $t$ refers to the training timesteps, $\lambda(t)= \text{sigmoid}(- k \cdot (t - t_0))$ is set to the sigmoid annealing function.

It should be noted in $\mathcal{L}_\text{PIT}$, we apply permutation invariant training (PIT) \cite{yu2017permutation, kolbaek2017multitalker} to remedy the source permutation problem, while the permutation in T-S $\mathcal{L}_\text{LTS}$ loss is determined by the teacher model.

With objective shifting, at the beginning of the training process, the student model is solely guided with the teacher's predictions, as soft label is believed to provide richer indication of teacher's behavior, thus leading to more efficient starting.
As training continues, the student gradually reduces the loss weight from the teacher, with more emphasis on clean reference, until the end of the training process, where the student completely learns from the clean target, to escape the limitation of the teacher's knowledge.

\subsection{Unlabeled Data training}
Training data for speech separation is generally artificially synthesized, so it requires clean speech as well as various noises. 
However, real overlapped data is slightly different from the artificially synthesized data, and it is hard to obtain the ground-truth of the unmixing results. The gap between artificial training data and the test data in the real scenario is a potential issue for speech separation. 
In this paper, we aim to leverage large-scale unlabeled mixing data in T-S learning.
In this way, the student model can approach the teacher model by mimicking the teacher’s behaviours, not only on the limited annotated data but also the large-scale unlabeled data.

Specifically, the student model is trained with T-S learning for two stages.
In the first stage, we pretrain the student model with the layer-wise T-S learning mechanism (Eq.~\ref{eq:lts}) on the large-scale unlabeled mixing data.
The student model learns to reproduce the final prediction and intermediate outputs of the teacher's model on the real overlapped data.
In the second stage, we train the student model with the layer-wise T-S learning and objective shifting (Eq.~\ref{eq:os}) on the annotated training data.
The student model begins with mimicking the teacher's behaviours on the annotated data, and ends with learning from the golden predictions of the annotated training data.

\section{Experiment}

\subsection{Datasets}
In this work, except the unlabeled learning part, all models are trained with 219 hours of artificially reverberated and mixed speech signals  sampled randomly from WSJ1 \cite{wsj1}. Following \cite{yoshioka2018multi}, we include four different mixture types in the training data. Each training mixture is generated by randomly picking one or two speakers from the WSJ1 dataset and convolving each with a 7 channel room impulse response (RIR) simulated with the image method \cite{habets2006room}. Then, we rescale and combine them with a source energy ratio between -5 and 5 dB. Simulated isotropic noise \cite{habets2007generating} is also added  at a 0--10 dB signal to noise ratio. The average overlap ratio of the training set is around 50\%. 
For the unlabeled data training, we apply the LibriVox and a Microsoft in-house dataset. LibriVox contains over 60k hours of audio derived from open-source audio books \cite{librilight}. The Microsoft in-house dataset contains 564 hours recording of discussion from Microsoft employees. 
We create 2k hours and 600 hours speech mixtures for LibriVox and the in-house dataset, by simply mixing two single speaker utterances. As in-house recording contains a noticeable amount of noise, there is no clean reference for mixtures derived from this dataset. 
We evaluate the models on the LibriCSS dataset \cite{chen2020continuous}, which consists of 10 hours of concatenated and mixed LibriSpeech utterances played and recorded in a meeting room. 
We test our model performance for both single channel and seven-channel setting, with word error rate(WER) as evaluation metric. We conducted both the utterance-wise evaluation and continuous input evaluation (refer to \cite{chen2020continuous} for the two evaluation schemes).

\subsection{Implementation Details}

The teacher model is the Conformer model from \cite{chen2020continuous2} which contains 16 encoder layers, 256 attention dimensions and 2048 FFN dimensions, resulting in 26.49M and 26.09M parameters for multi-channel and single-channel evaluation respectively. 
For multi-channel evaluation, the student Transformer model with 3.89M parameters consists of 6 encoder layers with 2 attention heads, 128 attention dimensions and 2048 FFN dimensions. 
The layer mapping function of the Layer-wise T-S learning is defined as $g(i) = \max(3 \times i - 2, 0)$.
For single-channel evaluation, the student Transformer model with 7.25M parameters consists of 12 encoder layers with 4 attention heads, 128 attention dimensions and 2048 FFN dimensions.  
The layer mapping function is defined as $g(i) = \min(2 \times i, i + 4)$.
The models are trained with the AdamW optimizer \cite{loshchilov2018decoupled} where the weight decay is set to 1e-2, the learning rate is 1e-4. We use the warm-up learning schedule with linear decay where the warm-up step is 10k, and the training step is 260k. 
For Object Shifting, we set $t_0$ to 150k, and select the best $k$ in \{1e-4, 5e-4\}.
For the unlabeled data training, we select the best $t_0$ in \{10k, 20K\}.
The small Transformer trained from scratch, denoted as Transformer-small$_{\text{Baseline}}$ is used as baseline system. 
The vanilla T-S learning, T-S learning with objective shifting, and layer-wise T-S learning are denoted as Transformer-small$_{\text{vanilla TS}}$, Transformer-small$_{\text{OS}}$, and Transformer-small$_{\text{LTS}}$ respectively.

We evaluate the speech separation accuracy with two ASR models. One is a hybrid system with a BLSTM based acoustic model and a 4-gram language model as used in the LibriCSS paper \cite{chen2020continuous}. The other is one of the best open source end-to-end Transformer \cite{wang2020semantic} based ASR models\footnote{https://github.com/MarkWuNLP/SemanticMask} which achieves WERs of 2.08\% and 4.95\% for LibriSpeech test-clean and test-other, respectively. 
We follow the sliding window-based CSS processing in continuous speech separation \cite{chen2020continuous} where the window size is set to 2.4s.
As with \cite{chen2020continuous}, we generate the individual target signals with spectral masking and mask-based adaptive minimum variance distortionless response (MVDR) beamforming for the single-channel and seven-channel cases,  respectively.

\subsection{Evaluation Results}
The result for utterance-wise and continuous separation are shown in 
Table \ref{tab:utt_result} and \ref{tab:con_result}.
We analyze the experiment results from three aspects: comparison with the teacher model, baseline small Transformer model, and models with unlabeled data training.

\begin{table*}[!t]
    \centering
    \footnotesize
    \caption{ Utterance-wise evaluation for seven-channel and single-channel settings. Two numbers in a cell denote \%WER of the \textbf{hybrid ASR model} used in LibriCSS \cite{chen2020continuous} and \textbf{E2E Transformer} based ASR model \cite{wang2020semantic}. 0S and 0L are utterances with short/long inter-utterance silence.
    }
    \label{tab:utt_result}

    \begin{tabular}{l|cccccc|c} 
         \toprule \hline
		\multirow{2}{*}{\textbf{System}} &
		\multicolumn{6}{c}{\textbf{Overlap ratio in \%}} & \multirow{2}{*}{\textbf{Avg gains}}  \\  & 0S & 0L & 10 & 20 & 30 & 40  \\  \hline
		    &   \multicolumn{6}{c}{Seven-channel Evaluation} \\ 
		\hline
		Conformer (Teacher)  & 7.0/3.1 & 7.2/3.2 & 8.9/3.6 & 11.1/4.6 & 13.6/5.8 & 15.1/6.3 & 13.2\%/15.4\% \\
		Transformer-small $_\text{Baseline}$  & 8.1/3.4 & 8.5/3.4 & 10.6/4.3 & 12.4/5.3 & 15.2/6.6 & 17.8/8.0 & 0.0\%/0.0\% \\
		\hline
		Transformer-small $_\text{vanilla TS}$   & 7.6/3.3 & 7.9/3.4 & 10.0/3.9 & 12.3/5.2 & 15.0/6.8 & 17.2/7.4 &  3.3\%/3.8\%  \\
		Transformer-small $_\text{LTS}$  & 7.3/3.3 & 7.7/3.2 & 9.6/4.0 & 12.2/5.1 & 14.8/6.8 & 17.2/7.6 & 5.0\%/3.8\% \\
		Transformer-small $_\text{OS}$  & 7.4/3.3 & 7.7/3.3 & 10.1/4.0 & 12.2/5.1 & 14.8/6.6 & 17.0/7.4 & 5.0\%/3.8\% \\
		Transformer-small $_\text{LTS +  OS}$  & 7.7/3.4 & 8.0/3.3 & 10.0/3.9 & 11.9/5.0 & 14.6/6.4 & 16.3/7.4 &  5.8\%/5.8\%\\
      Transformer-small $_\text{unlabeled LTS + OS}$  & 7.2/3.2 & 7.4/3.3 & 9.2/3.8 & 11.5/4.9 & 14.2/6.2 & 16.0/6.9 & 9.9\%/9.6\% \\					
		 \hline
        &   \multicolumn{6}{c}{Single-channel Evaluation}  \\ \hline
        
		Conformer (Teacher)  & 10.2/4.2 & 10.0/4.4 & 13.3/6.6 & 17.9/10.0 & 22.1/13.2 & 26.7/15.6 & 23.4\%/28.6\%  \\
		Transformer-small $_\text{Baseline}$  & 12.5/4.8 & 12.0/4.4 & 17.0/8.5 & 23.8/13.7 & 30.0/19.3 & 35.5/25.1 & 0.0\%/0.0\% \\
		\hline
		Transformer-small $_\text{vanilla TS}$  & 12.0/4.2 & 11.9/4.0 & 16.9/8.7 & 23.4/13.7 & 29.6/19.8 & 35.6/25.9 & 0.9\%/-0.8\% \\
		Transformer-small $_\text{LTS}$  & 12.0/4.1 & 11.8/4.1 & 16.7/8.6 & 22.6/13.8 & 29.0/19.7 & 35.0/25.3 &  2.8\%/0.0\% \\
		Transformer-small $_\text{OS}$  & 12.6/4.7 & 12.2/4.3 & 16.9/8.5 & 23.4/13.4 & 29.5/19.4 & 35.1/24.9 & 0.9\%/0.8\% \\
		Transformer-small $_\text{LTS +  OS}$  & 12.2/4.4 & 12.0/4.5 & 16.1/8.3 & 22.0/13.1 & 27.7/18.1 & 33.4/23.1 &  5.5\%/5.6\% \\					
        Transformer-small $_\text{unlabeled LTS + OS}$  & 11.0/4.3 & 10.8/4.5 & 15.0/7.6 & 20.6/11.6 & 25.6/16.3 & 31.1/20.2 & 12.8\%/14.3\% \\
		\hline
		\bottomrule
    \end{tabular}
\end{table*}

\begin{table*}[!t]
    \centering
    \footnotesize
    \caption{ Continuous speech separation evaluation for seven-channel and single-channel settings.}
    \label{tab:con_result}

    \begin{tabular}{l|cccccc|c} 
         \toprule \hline
		\multirow{2}{*}{\textbf{System}} &
		\multicolumn{6}{c}{\textbf{Overlap ratio in \%}} & \multirow{2}{*}{\textbf{Avg gains}}\\  & 0S & 0L & 10 & 20 & 30 & 40   \\ \hline
		    &   \multicolumn{6}{c}{Seven-channel Evaluation} \\ 
		\hline
		Conformer (Teacher)  & 11.8/5.7 & 9.0/4.1 & 13.2/6.3 & 14.1/7.1 & 18.6/9.8 & 20.3/10.8 & 9.9\%/18.9\% \\
		Transformer-small $_\text{Baseline}$  & 12.7/6.6 & 10.1/5.5 & 15.1/8.1 & 15.7/9.0 & 21.0/12.3 & 22.2/12.6 & 0.0\%/0.0\% \\
		\hline
		Transformer-small $_\text{LTS + OS}$  & 12.3/6.6 & 9.6/5.1 & 14.6/7.2 & 15.5/8.7 & 20.1/11.6 & 22.7/12.9 & 1.9\%/3.3\% \\
         Transformer-small $_\text{unlabeled LTS + OS}$  & 12.2/6.1 & 9.2/4.6 & 14.1/7.2 & 14.7/7.8 & 20.1/11.1 & 21.0/12.3 & 5.6\%/8.9\% \\					
		\hline
		    &   \multicolumn{6}{c}{Seven-channel Evaluation (w/o MVDR)} \\ 
		\hline
		Conformer (Teacher)  & 13.9/6.3 & 11.7/5.1 & 15.2/8.1 & 19.1/10.3 & 24.0/14.5 & 27.5/16.4 & 21.5\%/30.3\% \\
		Transformer-small $_\text{Baseline}$  & 18.0/9.4 & 15.3/8.5 & 20.5/11.6 & 24.4/14.9 & 30.0/19.1 & 33.9/23.2 & 0.0\%/0.0\% \\
		\hline
		Transformer-small $_\text{LTS + OS}$  & 16.4/8.9 & 14.0/8.1 & 18.2/10.4 & 22.5/13.7 & 27.4/18.1 & 32.0/21.8  & 8.0\%/6.9\% \\

        Transformer-small $_\text{unlabeled LTS + OS}$  & 14.0/7.5 & 12.2/6.5 & 16.0/9.5 & 19.6/12.1 & 24.9/16.6 & 29.0/19.7 & 18.6\%/17.2\% \\ 					
		\hline					
        &   \multicolumn{6}{c}{Single-channel Evaluation}  \\ \hline
        
		Conformer (Teacher)  & 16.4/9.6 & 15.0/9.0 & 19.3/12.1 & 24.3/15.6 & 29.1/20.5 & 32.4/23.5 & 36.1\%/49.0\% \\
		Transformer-small $_\text{Baseline}$  & 30.7/23.0 & 28.5/25.3 & 31.3/25.2 & 37.0/29.6 & 41.3/34.4 & 45.4/40.1 & 0.0\%/0.0\% \\
		\hline
		Transformer-small $_\text{LTS + OS}$  & 28.5/23.2 & 25.5/22.4 & 28.9/23.5 & 34.5/28.4 & 38.0/32.6 & 42.5/36.6 & 7.6\%/6.1\% \\ 
		Transformer-small $_\text{unlabeled LTS + OS}$  & 22.9/17.8 & 21.0/20.0 & 24.0/19.0 & 28.8/22.1 & 33.1/26.6 & 37.2/29.5 & 22.1\%/24.0\% \\					
		\bottomrule
    \end{tabular}
\end{table*}

\textbf{A comparison with the teacher model. }
Compared to the Conformer teacher model, the Transformer-small model with much less parameters can achieve an ultra faster speech separation speed.
We can obtain 21.5$\times$ and 11.4$\times$ speed-up for seven-channel and single-channel continuous speech separation with 2.4s window size.
Even if the runtime cost is largely reduced, we observe performance degradation in all experiments, but the seven channel degradation is not as serious as the single channel.  We guess the MVDR component bridges the gap between different models. To prove our hypothesis, we remove the MVDR in seven channel, and observe the gap between teacher and student becomes larger as shown in Table~\ref{tab:con_result}.

\textbf{A comparison with training from scratch.} We can achieve significant improvements with the proposed T-S learning method, compared to training from scratch, especially on the highly overlapped cases. 
For the seven-channel settings, we can obtain 5.8\% average relative WER gains with both the hybrid and E2E ASR systems for the utterance-wise evaluation. 
If we remove either the Layer-wise T-S Learning or Objective Shifting mechanism, performance drops are witnessed.
It shows that the student model can benefit from the intermediate knowledge from the teacher model and more knowledge from the training datasets.

For the single-channel settings, due to the limited input information, we experiment with the deeper Transformer-small model with more parameters.
Similar to the seven-channel cases, our T-S learning method can consistently outperform the baseline by a large margin, 
and achieve over 5\% relative WER gains for utterance-wise evaluation and over 6\% relative WER gains for continuous evaluation on average.

\textbf{Leveraging more unlabeled data.} 
By leveraging more unlabeled data, we can further boost the performance improvements of our proposed T-S learning methods.
For the single-channel settings, with the student model pretrained on the large-scale unlabeled data and shifted learning objective on the annotated training data, we can obtain 14.3\% and 24.0\% average relative WER gains for utterance-wise evaluation and continuous evaluation with E2E ASR systems.
For the seven-channel evaluation, utilizing more unlabeled data, we can obtain 9.6\% and 8.9\% average relative WER gains for utterance-wise evaluation and continuous evaluation with E2E ASR systems.
If we remove MVDR in seven-channel settings, our T-S learning methods can bring more significant improvements and  17.2\% average relative WER gains can be witnessed.

\section{Conclusions}

Because of the ultra fast inference speed, the small speech separation Transformer model is preferred for the deployment on devices.
In this work, we elaborate Teacher Student learning for better training of the ultra fast speech separation model.
The small student model is trained to reproduce the separation results of a large pretrained teacher model.
We also introduce Layer-wise Teacher Student Learning and Objective Shifting mechanisms to benefit the Teacher Student learning with more transferred knowledge.
The experimental results show the proposed methods can successfully improve the separation results of the small Transformer model.
Furthermore, pretraining on unlabeled data can further enhance the improvement.

\bibliographystyle{IEEEtran}

\bibliography{mybib}

\begin{thebibliography}{10}
\providecommand{\url}[1]{#1}
\csname url@samestyle\endcsname
\providecommand{\newblock}{\relax}
\providecommand{\bibinfo}[2]{#2}
\providecommand{\BIBentrySTDinterwordspacing}{\spaceskip=0pt\relax}
\providecommand{\BIBentryALTinterwordstretchfactor}{4}
\providecommand{\BIBentryALTinterwordspacing}{\spaceskip=\fontdimen2\font plus
\BIBentryALTinterwordstretchfactor\fontdimen3\font minus
  \fontdimen4\font\relax}
\providecommand{\BIBforeignlanguage}[2]{{%
\expandafter\ifx\csname l@#1\endcsname\relax
\typeout{** WARNING: IEEEtran.bst: No hyphenation pattern has been}%
\typeout{** loaded for the language `#1'. Using the pattern for}%
\typeout{** the default language instead.}%
\else
\language=\csname l@#1\endcsname
\fi
#2}}
\providecommand{\BIBdecl}{\relax}
\BIBdecl

\bibitem{dong2018speech}
L.~Dong, S.~Xu, and B.~Xu, ``Speech-transformer: a no-recurrence
  sequence-to-sequence model for speech recognition,'' in \emph{2018 IEEE
  International Conference on Acoustics, Speech and Signal Processing
  (ICASSP)}.\hskip 1em plus 0.5em minus 0.4em\relax IEEE, 2018, pp. 5884--5888.

\bibitem{Li2020Comparison}
J.~Li, Y.~Wu, Y.~Gaur, C.~Wang, R.~Zhao, and S.~Liu, ``On the comparison of
  popular end-to-end models for large scale speech recognition,'' in
  \emph{Proc. Interspeech}, 2020.

\bibitem{chang2020end}
X.~Chang, W.~Zhang, Y.~Qian, J.~Le~Roux, and S.~Watanabe, ``End-to-end
  multi-speaker speech recognition with transformer,'' in \emph{2020 IEEE
  International Conference on Acoustics, Speech and Signal Processing
  (ICASSP)}.\hskip 1em plus 0.5em minus 0.4em\relax IEEE, 2020, pp. 6134--6138.

\bibitem{chen2020dual}
J.~Chen, Q.~Mao, and D.~Liu, ``Dual-path transformer network: Direct
  context-aware modeling for end-to-end monaural speech separation,''
  \emph{Proc. Interspeech 2020}, pp. 2642--2646, 2020.

\bibitem{chen2020continuous2}
S.~Chen, Y.~Wu, Z.~Chen, J.~Li, C.~Wang, S.~Liu, and M.~Zhou, ``Continuous
  speech separation with conformer,'' \emph{arXiv preprint arXiv:2008.05773},
  2020.

\bibitem{li2014learning}
J.~Li, R.~Zhao, J.-T. Huang, and Y.~Gong, ``Learning small-size {DNN} with
  output-distribution-based criteria.'' in \emph{Proc. Interspeech}, 2014, pp.
  1910--1914.

\bibitem{hinton2015distilling}
G.~Hinton, O.~Vinyals, and J.~Dean, ``Distilling the knowledge in a neural
  network,'' \emph{arXiv preprint arXiv:1503.02531}, 2015.

\bibitem{watanabe2017student}
S.~Watanabe, T.~Hori, J.~Le~Roux, and J.~R. Hershey, ``Student-teacher network
  learning with enhanced features,'' in \emph{2017 IEEE International
  Conference on Acoustics, Speech and Signal Processing (ICASSP)}.\hskip 1em
  plus 0.5em minus 0.4em\relax IEEE, 2017, pp. 5275--5279.

\bibitem{subramanian2018student}
A.~S. Subramanian, S.-J. Chen, and S.~Watanabe, ``Student-teacher learning for
  blstm mask-based speech enhancement,'' \emph{Proc. Interspeech 2018}, pp.
  3249--3253, 2018.

\bibitem{chen2020continuous}
Z.~Chen, T.~Yoshioka, L.~Lu, T.~Zhou, Z.~Meng, Y.~Luo, J.~Wu, X.~Xiao, and
  J.~Li, ``Continuous speech separation: Dataset and analysis,'' in \emph{2020
  IEEE International Conference on Acoustics, Speech and Signal Processing
  (ICASSP)}.\hskip 1em plus 0.5em minus 0.4em\relax IEEE, 2020, pp. 7284--7288.

\bibitem{wang2014training}
Y.~Wang, A.~Narayanan, and D.~Wang, ``On training targets for supervised speech
  separation,'' \emph{IEEE/ACM Transactions on Audio, Speech, and Language
  Processing}, vol.~22, no.~12, pp. 1849--1858, 2014.

\bibitem{erdogan2017deep}
H.~Erdogan, J.~R. Hershey, S.~Watanabe, and J.~Le~Roux, ``Deep recurrent
  networks for separation and recognition of single-channel speech in
  nonstationary background audio,'' in \emph{New Era for Robust Speech
  Recognition}.\hskip 1em plus 0.5em minus 0.4em\relax Springer, 2017, pp.
  165--186.

\bibitem{vaswani2017attention}
A.~Vaswani, N.~Shazeer, N.~Parmar, J.~Uszkoreit, L.~Jones, A.~N. Gomez,
  {\L}.~Kaiser, and I.~Polosukhin, ``Attention is all you need,'' in
  \emph{Advances in neural information processing systems}, 2017, pp.
  5998--6008.

\bibitem{shaw2018self}
P.~Shaw, J.~Uszkoreit, and A.~Vaswani, ``Self-attention with relative position
  representations,'' in \emph{Proceedings of the North American Chapter of the
  Association for Computational Linguistics}, 2018, pp. 464--468.

\bibitem{chen2020don}
S.~Chen, Y.~Wu, Z.~Chen, T.~Yoshioka, S.~Liu, and J.~Li, ``Don't shoot
  butterfly with rifles: Multi-channel continuous speech separation with early
  exit transformer,'' \emph{arXiv preprint arXiv:2010.12180}, 2020.

\bibitem{sun2019patient}
S.~Sun, Y.~Cheng, Z.~Gan, and J.~Liu, ``Patient knowledge distillation for bert
  model compression,'' in \emph{Proceedings of the 2019 Conference on Empirical
  Methods in Natural Language Processing and the 9th International Joint
  Conference on Natural Language Processing (EMNLP-IJCNLP)}, 2019, pp.
  4314--4323.

\bibitem{kiperwasser2018scheduled}
E.~Kiperwasser and M.~Ballesteros, ``Scheduled multi-task learning: From syntax
  to translation,'' \emph{Transactions of the Association for Computational
  Linguistics}, vol.~6, pp. 225--240, 2018.

\bibitem{chen2020recall}
S.~Chen, Y.~Hou, Y.~Cui, W.~Che, T.~Liu, and X.~Yu, ``Recall and learn:
  Fine-tuning deep pretrained language models with less forgetting,'' in
  \emph{Proceedings of the 2020 Conference on Empirical Methods in Natural
  Language Processing (EMNLP)}, 2020, pp. 7870--7881.

\bibitem{yu2017permutation}
D.~Yu, M.~Kolb{\ae}k, Z.-H. Tan, and J.~Jensen, ``Permutation invariant
  training of deep models for speaker-independent multi-talker speech
  separation,'' in \emph{2017 IEEE International Conference on Acoustics,
  Speech and Signal Processing (ICASSP)}.\hskip 1em plus 0.5em minus
  0.4em\relax IEEE, 2017, pp. 241--245.

\bibitem{kolbaek2017multitalker}
M.~Kolb{\ae}k, D.~Yu, Z.-H. Tan, and J.~Jensen, ``Multitalker speech separation
  with utterance-level permutation invariant training of deep recurrent neural
  networks,'' \emph{IEEE/ACM Transactions on Audio, Speech, and Language
  Processing}, vol.~25, no.~10, pp. 1901--1913, 2017.

\bibitem{wsj1}
L.~D.~C. Philadelphia, ``{CSR-II (WSJ1) Complete},'' 1994,
  \url{http://catalog.ldc.upenn.edu/LDC94S13A}.

\bibitem{yoshioka2018multi}
T.~Yoshioka, H.~Erdogan, Z.~Chen, and F.~Alleva, ``Multi-microphone neural
  speech separation for far-field multi-talker speech recognition,'' in
  \emph{2018 IEEE International Conference on Acoustics, Speech and Signal
  Processing (ICASSP)}.\hskip 1em plus 0.5em minus 0.4em\relax IEEE, 2018, pp.
  5739--5743.

\bibitem{habets2006room}
E.~A. Habets, ``Room impulse response generator,'' \emph{Technische
  Universiteit Eindhoven, Tech. Rep}, vol.~2, no. 2.4, p.~1, 2006.

\bibitem{habets2007generating}
E.~A. Habets and S.~Gannot, ``Generating sensor signals in isotropic noise
  fields,'' \emph{The Journal of the Acoustical Society of America}, vol. 122,
  no.~6, pp. 3464--3470, 2007.

\bibitem{librilight}
J.~{Kahn}, M.~{Rivière}, W.~{Zheng}, E.~{Kharitonov}, Q.~{Xu}, P.~E.
  {Mazaré}, J.~{Karadayi}, V.~{Liptchinsky}, R.~{Collobert}, C.~{Fuegen},
  T.~{Likhomanenko}, G.~{Synnaeve}, A.~{Joulin}, A.~{Mohamed}, and E.~{Dupoux},
  ``Libri-light: A benchmark for asr with limited or no supervision,'' in
  \emph{2020 IEEE International Conference on Acoustics, Speech and Signal
  Processing (ICASSP)}, 2020, pp. 7669--7673,
  \url{https://github.com/facebookresearch/libri-light}.

\bibitem{loshchilov2018decoupled}
I.~Loshchilov and F.~Hutter, ``Decoupled weight decay regularization,'' in
  \emph{International Conference on Learning Representations}, 2018.

\bibitem{wang2020semantic}
C.~Wang, Y.~Wu, Y.~Du, J.~Li, S.~Liu, L.~Lu, S.~Ren, G.~Ye, S.~Zhao, and
  M.~Zhou, ``Semantic mask for transformer based end-to-end speech
  recognition,'' \emph{Proc. Interspeech 2020}, pp. 971--975, 2020.

\end{thebibliography}

\end{document}